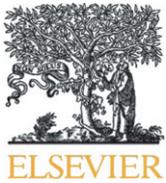
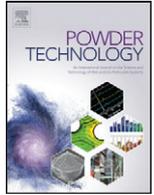

# Bouncing behavior and dissipative characterization of a chain-filled granular damper

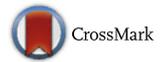

Cheng Xu [a], Ning Zheng [a,b,\*], Liang-sheng Li [c], Qing-fan Shi [a,\*\*]

[a] School of Physics, Beijing Institute of Technology, Beijing 100081, China
[b] Key Laboratory of Cluster Science of Ministry of Education, Beijing 100081, China
[c] Science and Technology on Electromagnetic Scattering Laboratory, Beijing 100854, China

## ABSTRACT

We have experimentally investigated the bouncing behavior and damping performance of a container partially filled with granular chains, namely a chain-filled damper. The motion of the chain-filled damper, recorded by a particle tracing technology, demonstrates that the granular chains can efficiently absorb the collisional energy of the damper. We extract both the restitution coefficient of the first collision and the total flight time to characterize the dissipation ability of the damper. Two containers and three types of granular chains, different in size, stiffness and restitution coefficient, are used to examine the experimental results. We find that the restitution coefficient of the first collision of a single-chain-filled damper can linearly tend to vanish with increasing the chain length and obtain a minimum filling mass required to cease the container at the first collision (no rebound). When the strong impact occurs, the collisional absorption efficiency of a chain-filled damper is superior to a monodisperse-particle-filled damper. Furthermore, the longer the chains are, the better the dissipative effect is.

© 2016 Elsevier B.V. All rights reserved.

## 1. Introduction

In the field of civil engineering, machinery manufacturing, and aerospace industry, mechanical vibration often results in deterioration and even failure of equipment performance [1]. To address this issue, a vibration damper has been introduced to minimize the irritating vibration in a wide variety of applications, including vibration attenuation of metal cutting machines [2], turbine blade oscillation [3], vibrating antennae [4], sports equipment [5], medical tools [6], etc. In contrast with a conventional damper, a granular damper constitutes a promising technology for designing a next-generation damper, because it is efficient over a range of driven-frequencies, low-cost, long lifetime, easy maintenance and insensitive to external temperature [7–9]. The granular damper is a container partly filled with granular particles, which is attached to or embedded in a vibrating structure to weaken the external vibration. During the vibration, the energy or momentum from the vibrating structure is transferred to filling particles. Then the transferred energy is rapidly dissipated through the inter-particle friction and inelastic inter-particle and particle-wall collisions in finite time. As a result, the vibration is considerably attenuated or even eliminated.

The dissipation efficiency or performance of a granular damper relies on many system parameters, such as the geometry of the container, material properties of filling particles, and external vibrations. For example, the damping performance between a piston-type and a box-type granular damper was compared by a numerical simulation [10]. It was found that for the box-type damper, a uniform energy dissipation occurred because almost all particles contributed equally to the energy transfer. In contrast, for the piston-type damper, only local particles directly under the piston were involved in the energy transfer. In general, the box-type damper is more dissipation efficient than the piston-type damper only due to the geometry. As stated before, both the friction and collision contribute to the energy dissipation during a damping process. The collisional intensity and frequency of inter-particle collisions significantly depend on the fluidization extent of granular particles. Thus, collision contributes more to energy dissipation at low volume fraction; friction contributes more at low vibration frequency [11]. Furthermore, the relative significance between the friction and collision also depends strongly on the particle size [12]. The friction often plays a more important role in damping than the collision for small particles, but the collision becomes dominant as the particle size increases. These examples above show that for various situations an appropriate damper ought to be carefully chosen to achieve the optimal dissipation. It is therefore necessary to study the dissipation performance of a variety of granular dampers at different conditions, which may be potentially helpful for the engineering.

Almost most of previous efforts, if not all, focus on the granular damper filled with monodisperse particles. It is very attractive to compare the performance of energy dissipation between monodisperse particle-filled dampers and other dampers with different filling materials, which is potentially useful to design new granular damping systems. Compared with the monodisperse particle, the granular chain, mainly acting as an experimental analogy with molecular chains in most cases [13–15], is a promising candidate to provide competitive dissipation performance. Although considerable researches involving granular chains have been performed, the mechanical behavior of granular chains in many aspects did not receive enough attention yet [16]. The investigation on the chain-filled damper appears to be absent. The connection between the dissipation efficiency and the material properties of the damper needs to be explored in details.

\* Corresponding author at: Key Laboratory of Cluster Science of Ministry of Education, Beijing 100081, China.
\*\* Corresponding author.
E-mail addresses: Ningzheng@bit.edu.cn (N. Zheng), liliangsheng@gmail.com (L. Li), qfshi123@bit.edu.cn (Q. Shi).

In the manuscript, we experimentally study the dynamics of a chain-filled damper which bounces on a flat plate. Two kinds of chain-filled dampers, namely the single-chain-filled and multiple-chains-filled damper, are used to measure the restitution coefficient of the first impact $\varepsilon_1$ and total flight time $\tau$ under different conditions. The relationships between the measured quantities and the external parameters such as filling mass, chain length, and clearance length are presented. On the basis of the measurement, we qualitatively compare the dissipation efficiency of a chain-filled damper with a monodisperse-particle-filled damper. In addition, we rescale the $\varepsilon_1$ for a single-chain-filled damper, and find that these curves collapse together in a linear fashion in which the underlying physics can be explained by using a momentum exchange model.

## 2. Experimental setup

The experimental setup is demonstrated in Fig. 1. The damper consists of a cylindrical, empty container and granular chains filled into the container. The inner diameter of the container is 30 mm, and the outer diameter is 40 mm. The low end of the container is closed with a rounded acrylic cap. The upper end is sealed by a baffle so that the clearance length of the container, $L$, can be changed by adjusting the position of the baffle. The case $L = \infty$ corresponds to the absence of the baffle. The semi-rigid chains inside the container are composed of hollow, steel balls and steel rods, which are similar with those used in previous work [14,17]. The rod, as a link connecting balls, is not fixed, but rather retractile. In order to verify the generality and robustness of the measurement results, different containers and granular chains are employed in the experiment. The initial status of the granular chains inside the container is always set to be random packing to minimize the effect from different initial packing status. The stationary container is released from a given height, and bounces after it collides with a massive steel base below. To ensure the rebound stability of the bouncing container, a glass tube with a slightly larger diameter is used to align the container vertically, and a heavy base with smooth, flat surface is placed on a cushion. With using a free-fall experiment to examine the air drag on the motion of the damper, it is found that the air drag appears to be negligible. A high-speed camcorder (Phantom V7.3) tracks the trajectories of the bouncing container in real time, and thus a wealth of information such as restitution coefficient can be accurately extracted by an imaging algorithm. To confirm the generality of the experiment results, two different containers A1 and A2 (the geometries of two containers A1 and A2 are same), and three types of chains C1, C2 and C3 are used (see details in Table 1 for their material properties). Unless otherwise noted, each measurement was repeated 5 times.

## 3. Results and discussion

### 3.1. Single-chain-filled damper

In this section only one single chain is placed into the container, and the filling mass $M_{fill}$ is proportional to the chain length $N$, $M_{fill} = Nm_\rho$, where $m_\rho$ is the mass per unit length of a chain. Unless specified otherwise, the upper end of a container is always free, namely $L = \infty$. Fig. 2(a) shows that the vertical position $h$ of the geometric center of a container is plotted as a function of time for different filling masses, or chain lengths. The first rebound height decreases as the chain length (the filling mass) increases. The restitution coefficient $\varepsilon = -V_a/V_b$ is displayed in Fig. 2(b) for the first five impacts, numbered by $n_b$, where $V_a$ and $V_b$ are the velocities of the container after and before the impact, respectively. For the first impact, the restitution coefficient $\varepsilon_1$ drops with the increase of the chain length, namely the filling mass. For the damper filled with short chains, $\varepsilon_1$ slightly decreases, and the chains hardly influence the dynamics of the container compared with an empty one. Specifically, for the empty container (N = 0) or the container with small filling masses (N = 16, 32), the restitution coefficient $\varepsilon$ is insensitive to $n_b$ within the measurement precision. This is because the energy transferred to the tiny filling mass is very small proportion of the total energy of the damping system. For the damper filled with long chains (N = 128, 256), $\varepsilon$ exhibits a complicated behavior with $n_b$. For instance, $\varepsilon_2$ rises instead. The fact can be understood that after the initial impact, the part of or the whole of the chain moves upward faster than the container, and forms a cluster at the top of or out of the container. When the second impact occurs, the container will be at a status of "free of chain" because the chain is in a status of free fall. For the case N = 512, $\varepsilon_2 = 0$ suggests that the part of the long chain possibly adheres to the container after the first impact, and damps the second rebound.

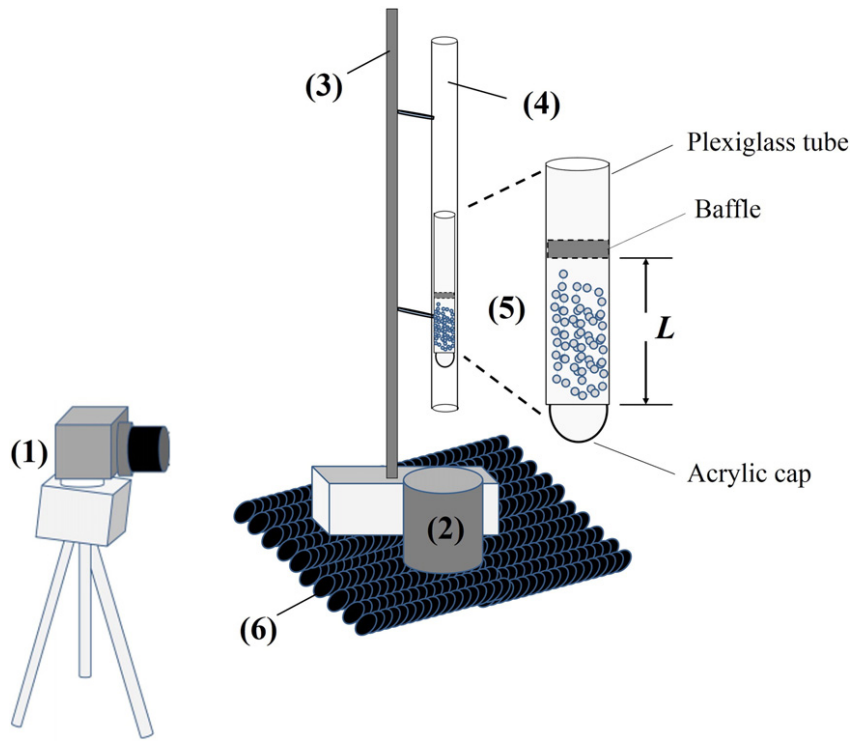

**Fig. 1.** Schematic of the experimental setup, not to scale. (1) High-speed camcorder (2) steel base, (3) mounting bracket, (4) glass tube, (5) plexiglass container, (6) cushion. The enlarged portion of the plexiglass container shows the basic structure, the description of which is detailed in the text.

**Table 1**
Properties of granular chains and containers used in the experiment.

| Chain | Diameter of sphere, a (mm) | Length of link, b (mm) | Mass per unit length, $m_\rho$ (g) | Minimum sphere number to form a loop |
|---|---|---|---|---|
| C1 | 3.00 ± 0.03 | 1.50 ± 0.03 | 0.061 ± 0.03 | 8 |
| C2 | 1.50 ± 0.03 | 0.50 ± 0.03 | 0.0096 ± 0.0005 | 10 |
| C3 | 6.00 ± 0.03 | 3.00 ± 0.03 | 0.324 ± 0.004 | 7 |

| Container | Mass, $M_c$ (g) (empty) | Restitution coefficient, $\varepsilon_0$ (empty) | Clearance length, L (mm) | Initial height $H_0$ (mm) |
|---|---|---|---|---|
| A1 | 37.9 ± 0.2 | 0.90 ± 0.03 | 20, 30, 40, 50, 60, 70, ∞ | 40, 50, 60, 70, 80, 90, 100 |
| A2 | 22.4 ± 0.2 | 0.80 ± 0.03 | ∞ | 100 |

Fig. 2(c) exhibits the chain length N dependence of the restitution coefficient of the first impact $\varepsilon_1$. It clearly shows that $\varepsilon_1$ in all container-chain combinations monotonically reduces to zero with the increase of the chain length, except for some range where $\varepsilon_1$ restarts to augment. If the x axis in Fig. 2(c) is converted into the filling mass by multiplying the mass per unit length from each type of chain, the curves in Fig. 2(c) converge together to form two clusters, as shown in Fig. 2(d). It implies that all data points obtained from a given container will collapse to one single curve regardless of the chains filled in this container. In Fig. 2(d) it is also found that $\varepsilon_1$ linearly decreases with the increase of the filling mass, and vanishes at some critical mass. Over the critical mass, $\varepsilon_1$ starts to augment once again. The abnormal behavior presumably relates to the speculation that the container and a part of the chain near the bottom of the container act as an integral solid, and the part of the chain thus no longer serves as the recipient of the momentum transferred from the container [18]. The immobile part effectively increases the mass of the container, resulting in the increase of the restitution coefficient.

The rescaled curves $\varepsilon_1 - \varepsilon_0$ in Fig. 2(e) can collapse together in a linear fashion within the range $0 < m_\rho N/M_c < 0.6$, where the $M_c$ is the mass of an empty container. All data points within this range are well described by the linear equation, $\varepsilon_1 - \varepsilon_0 = k\frac{m_\rho N}{M_c}$, and k = −1.28, independent of the material properties of the filling chains. It suggests that only the filling mass is the dominant factor for a single-chain-filled damper. The underlying physics in the linear expression can be explained by the three-stage momentum transfer model [19,20]. Note that the critical mass of the single-chain-filled damper (~$0.8\varepsilon_0 M_c$), namely the minimum mass required to cease the container at the first collision, is twice smaller than that of a monodisperse-particle-filled damper (~$1.5\varepsilon_0 M_c$) in previous report [19]. For a given container, the same dissipation can be obtained from a chain-filled damper with only half the filling mass. Therefore granular chains as filling materials are superior to monodisperse particles in the dissipation efficiency.

Total flight time $\tau$ is another relevant parameter which denotes the duration from initial free fall to a complete stop of the container. Apparently, the shorter the total flight time is, the higher the dissipation efficiency of a damper owns. Thus it can qualitatively characterize the dissipation efficiency of a damper. Fig. 2(f) shows the total flight time as a function of the chain length when the initial height is given. With the increase of the chain length, $\tau$ monotonically decreases for all combinations. All points in Fig. 2(f) cluster together to form a single scaling curve, as shown in Fig. 2(g). Here $\tau_0$ is the total flight time from an empty container, $\tau^*$ is the time between the initial free fall and first rebound. The best fitting curve can be described by an equation $(1 - Nm_\rho/M_{crit})/(1 + \alpha\frac{Nm_\rho}{M_{crit}})$ where the critical mass $M_{crit} \approx 0.8\varepsilon_0 M_c$, and $\alpha$ is a fitting parameter. The mathematical representation is similar with that from a monodisperse particle-filled damper.

### 3.2. Multiple-chains-filled damper

In this section, more than one single chain is filled in a container that is released from a given height $H_0 = 100$ mm. The number of chains s is inversely proportional to the chain length with a given filling mass, namely $s = \frac{M_{fill}}{Nm_\rho}$. In Fig. 3(a), it is found that the restitution coefficient of the first impact $\varepsilon_1$ decreases monotonically with the increase of the chain length N when the filling mass is given. It suggests that long chains acquire more energy compared with short chains during the first impact. Additionally, it also shows that $\varepsilon_1$ reduces with the increase of the filling mass if the chain length is fixed in Fig. 3(a). The filling mass dependence of $\varepsilon_1$ persists for all chain lengths. However, the relationship $\varepsilon_1 - \varepsilon_0 = k\frac{Nm_\rho}{M_c}$ is invalid for the multiple chains damper any more.

To examine the generality of the result above, the chains $C_2$ and $C_3$ and the container $A_2$ are used with a filling mass $M_{fill} = 0.2M_c$ arbitrarily chosen. The relation between $\varepsilon_1$ and N is similar with that in the Fig. 3(a), see Fig. 3(b). Fig. 3(c) shows the chain length dependence of the total flight time $\tau$ for different filling masses. The total flight time monotonically reduces with the chain length with a given filling mass. In other words, the container may stop more rapidly by filling long chains. This result confirms that long chains own higher dissipative efficiency. On the other hand, the $\tau$ decreases with the filling mass if the chain length is given, which is consistent with the result from the monodisperse-particle-filled damper. Similarly, Fig. 3(d) shows the total flight time as a function of the chain length with a given filling mass $M_{fill} = 0.2M_c$. With the increase of the chain length, $\tau$ monotonically deceases for all combinations and the robustness of the result is reconfirmed.

The initial height of the damper determines the total system energy. In Fig. 4(a), all restitution coefficients of the first impact $\varepsilon_1$ almost overlap with a given mass, which suggests that the dynamics of the first collision is independent of the initial height. Fig. 4(b) describes the chain length dependence of the total flight time $\tau$ at different initial heights. It is unexpectedly found that the initial height radically influences the dependence. At $H_0 = 20$ cm and 10 cm, $\tau$ falls with the increase of the chain length, suggesting that long chains have a strong dissipative efficiency. At $H_0 = 8$ cm, $\tau$ almost remains constant, and is insensitive to the chain length. At $H_0 = 5$ cm, the situation is contrary to the large height; namely the total flight time increases with the chain length. We also tested other chains and container, and obtained the similar result. It seems that for the strong impact, the long chains dissipate the energy more efficiently; and for the weak impact, the short chains have an advantage. However, the underlying physics of the height dependence of the dissipation efficiency still remains open.

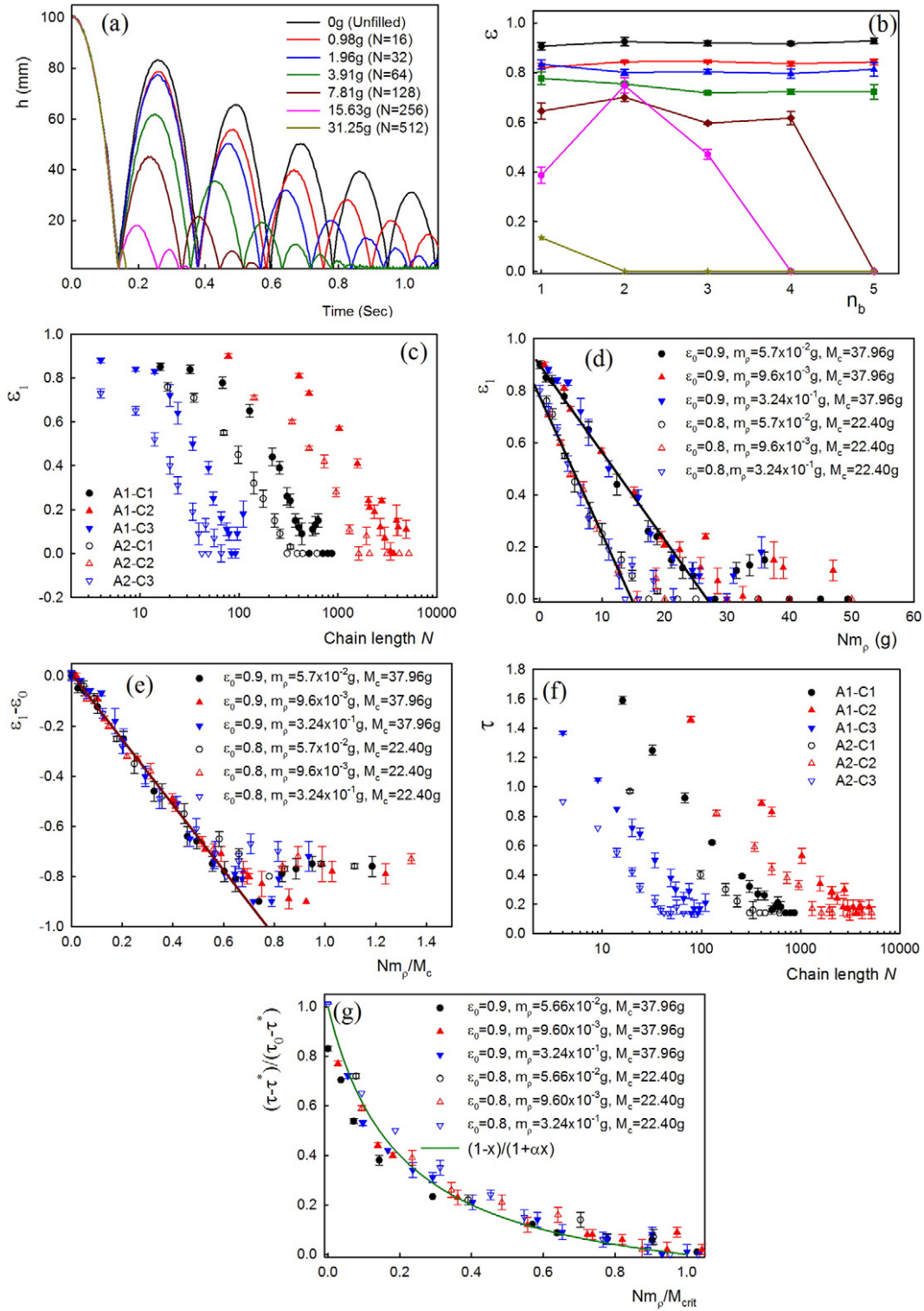

Fig. 2. Single-chain-filled damper. (a) Vertical position of the container as a function of time for different filling masses (chain lengths), where the chain C1 and the container A1 are used. (b) Restitution coefficient of the first five consecutive impacts, same chains and container as (a). Colors are in correspondence with (a). (c) Restitution coefficient $\varepsilon_1$ for the first impact vs chain length $N$ for different combinations using a semilogarithmic scale. A1: container 1, A2: container 2. C1: chain 1; C2: chain 2; C3: chain 3. For example, A1–C1 indicates chain 1 in the container 1. The specific parameters for these containers and chains are shown in the Table 1. (d) Restitution coefficient $\varepsilon_1$ vs $Nm_p$ from the data in (c), the solid lines are plotted for eye guide. (e) $\varepsilon_1 - \varepsilon_0$ vs $Nm_p/M_c$ from the data in (c). The points collapse together at the interval [0, 0.6]. (f) Total flight time $\tau$ vs chain length $N$ using a semilogarithmic scale. (g) $(\tau - \tau^*)/(\tau_0 - \tau^*)$ vs $Nm_p/M_c$ from all data in (f). All points collapse together to form a single curve, The best fitting curve can be described by the equation $(1 - Nm_p/M_{crit})/(1 + \alpha \frac{Nm_p}{M_{crit}})$, where the critical mass $M_{crit} \approx 0.8 \varepsilon_0 M_c$, and $\alpha = 5$.

### 3.3. Clearance length dependence

In previous studies, the baffle as the upper end of the container is always absent and the granular chains can even jump out of the container after strong impact occurs. The energy exchange between the container and chains only takes place as the chains impact the bottom or inner wall of the container. In this section, a baffle restricts the free room above the chain pile and increases collisional chances, which significantly influences the dynamics of the damper. The free room is characterized by the clearance length $L$ that is the distance from the baffle to the bottom of the container. In the experiment, the initial height $H_0 = 10$ cm is chosen such that the chains can at least collide with the baffle once for all $L$. With a given filling mass $M_{fill} = 15.6$ g, the Fig. 5(a) shows the height position of a single-chain-filled damper as a function of time for different $L$. Compared with the trajectory without the baffle, the presence of the upper boundary changes the dynamics of the damper, and the total flight time is decreased due to the energy dissipation from the upper boundary, see Fig. 5(b). Interestingly, the first rebound height increases with the decrease of $L$. This difference in the first rebound height is not attributed to the restitution coefficient of the container, which is the same for the first collision. The reason is that in the case with small $L$ more chains can hit the upper boundary and transfer their energy back to the container. The transfer augments the velocity of the container, and thus it can rise higher. In addition, for the damper with small $L$, the chains have no much free space to dissipate the same energy as that with large $L$, and therefore the container rebounds higher.

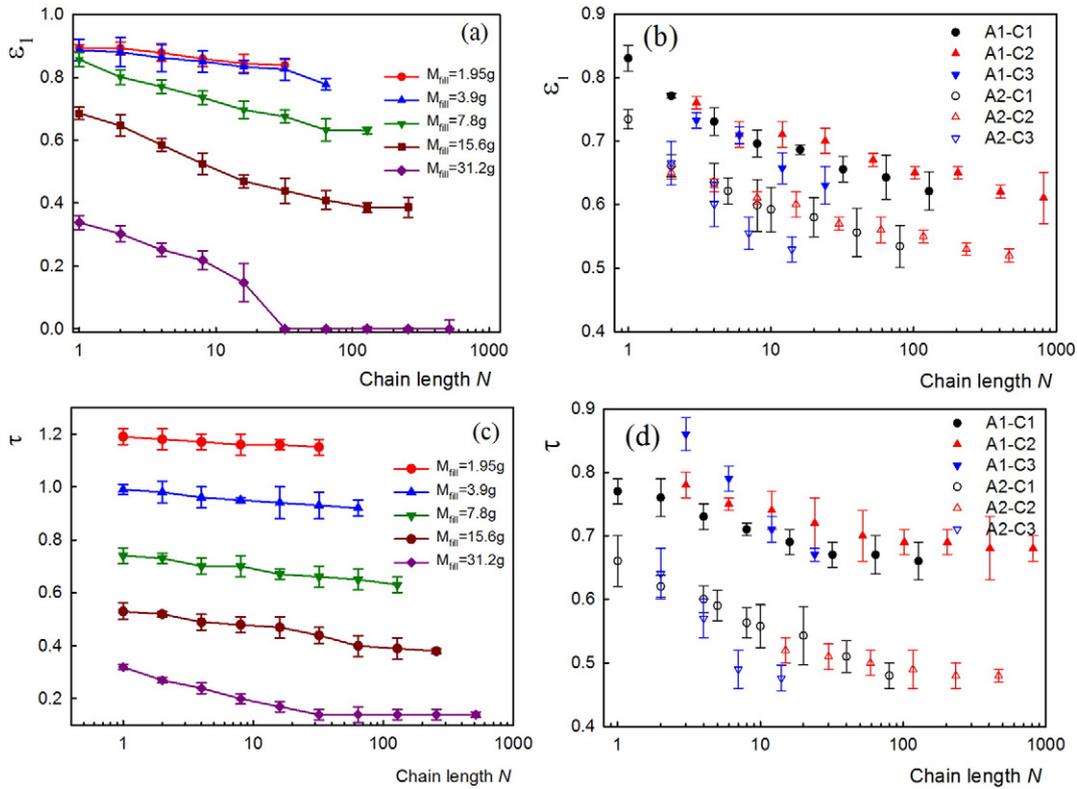

**Fig. 3.** Multiple-chains-filled damper. (a) Restitution coefficient of the first impact $\varepsilon_1$ vs chain length $N$ for different filling masses using a semilogarithmic scale, where the chain C1 and the container A1 are used. The number of chains $s$ is inversely proportional to the chain length with a given filling mass, $s = \frac{M_{fill}}{Nm_p}$. For example, the number $s = 128$ to 1 corresponding to the chain length $N = 1$–128 with a given filling mass $M_{fill} = 7.8$ g. (b) Restitution coefficient $\varepsilon_1$ of the first impact vs chain length $N$ for different combinations. The filling mass is 7.8 g in container A1 and 4.48 g in container A2, respectively. The range of chains number $s$ for A1–C1: 1–128; A1–C2: −271; A1–C3: 1–8; A2–C1: 1–80; A2–C2: 1–234; A2–C3: 1–7. (c) Total flight time $\tau$ vs chain length $N$ for different filling masses, same chains and container as (a). (d) Total flight time $\tau$ vs chain length $N$ for different combinations, same chains and container as (b).

The chain length dependence of the total flight time for different $L$ is also measured with a filling mass $M_{fill} = 7.8$ g arbitrarily given, as shown in the Fig. 5(b). It is found that the total flight time at $L = \infty$ is longer than other $L$, which is due to the confinement that dissipates more energy during the transfer of the momentum by the collision. Remarkably, All total flight time is weakly dependent on the chain length except for $L = \infty$, which suggests the chains can only interact with upper boundary once and dissipate most of the initial energy of the system, and during the following rebound the chains just collide the container bottom. Remarkably, the number $s$ of the chains ranges from 32 to 2 corresponding to chain length $N$ from 4 to 64. It suggests that the confinement effect of the upper boundary can eliminate the difference in the damping performance between the single-chain-filled and multiple-chains-filled damper since the total flight time is almost the same for all chain lengths.

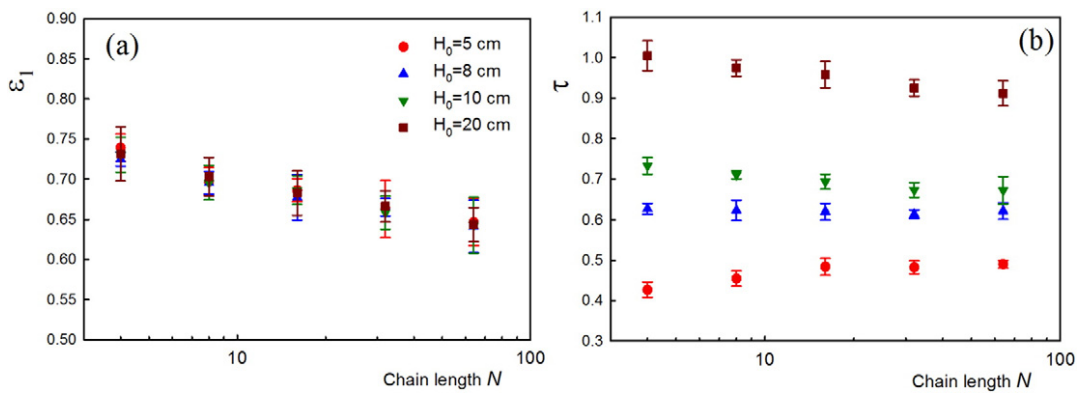

**Fig. 4.** The container A1 and the chain C1 are used to measure (a) Restitution coefficient $\varepsilon_1$ vs chain length $N$ for different initial heights, with a given filling mass. (b) Total flight time $\tau$ vs chain length $N$ for different initial heights, with a given filling mass. Colors are in correspondence with (a). The number $s$ of the chains in (a) and (b) ranges from 32 to 2 corresponding to chain length $N$ from 4 to 64, where $sN = 128$.

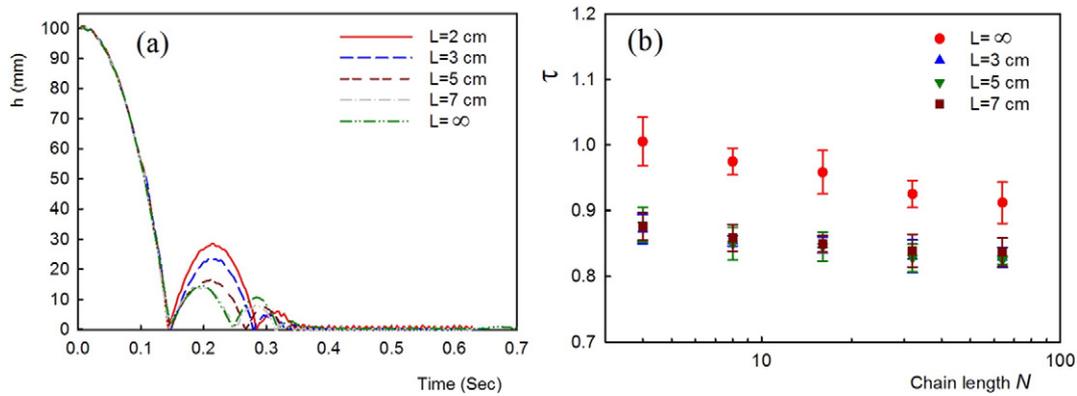

**Fig. 5.** The container A1 and the chain C1 are used to measure (a) Vertical position vs chain length $N$ for different clearance lengths $L$, with a given filling mass, a single chain with chain length $N = 256$ inside the container. (b) Total flight time $\tau$ vs chain length $N$ for different clearance lengths $L$, with a given filling mass. The number $s$ of the chains ranges from 32 to 2 corresponding to chain length $N$ from 4 to 64, where $sN = 128$.

## 4. Summary

In conclusion, we have experimentally investigated the bouncing behavior and damping properties of a chains-filled granular damper. We focus on the measurement of the restitution coefficient of the first impact $\varepsilon_1$ and total flight time $\tau$. $\varepsilon_1$ characterizes the energy partition between the container and granular chains after the first impact. We have studied the $\varepsilon_1$ for two types of dampers: single-chain-filled and multiple-chains-filled damper. For the single-chain-filled damper, its dynamics is similar with particle-filled damper. The rescaled curves $\varepsilon_1 - \varepsilon_0$ vs $M_{fill}$ collapse together to form a linear master curve, which indicates that only the filling mass $M_{fill}$ is the crucial factor within the studied range. We apply a simple model of momentum conservation to explain the linear relationship. For the multiple-chains-filled damper, the situation is much more complex. We find that $\varepsilon_1$ decreases with the chain length or filling mass when either of them is fixed. However, the linear relationship shown in the single-chain-filled damper is invalid any more.

Another important parameter, total flight time $\tau$, can qualitatively characterize the dissipation efficiency of a damper. In the case of the single-chain-filled damper, $\tau$ reduces with the chain length, and can be rescaled to form a single curve that may be described by a power law equation. Similarly, the decreasing dependence of $\tau$ still holds for the multiple-chains-filled damper, but the power law relationship is inapplicable.

We also compare the damping performance with various dampers, and find that the dissipation efficiency of the chains-filled damper is superior to the particle-filled damper when the strong impact occurs. Furthermore, the longer the chains are, the better the dissipative effect is. In contrast, the particle-filled damper performs better for the weak impact. Finally, we preliminarily study the effect of the clearance length on the dynamics of the damper, and discuss the experimental results.

In the future, if the dissipative mechanism of inter-chains interaction can be established, more light will be shed on the further understanding of dissipative performance of the chain-filled granular damper.

*Nomenclature*

| | |
|---|---|
| $a$ | Diameter of sphere on a chain |
| $b$ | Length of a link connecting two spheres |
| $N$ | Chain length |
| $s$ | Number of chains in a container |
| $m_\rho$ | Mass per unit length of granular chains |
| $M_{fill}$ | Filling mass of granular chains |
| $M_{crit}$ | Minimum mass required to cease the container at the first collision |
| $L$ | Clearance length |
| $M_c$ | Mass of an empty container |
| $\varepsilon_0$ | Restitution coefficient of an empty container |
| $\varepsilon_i$ | Effective restitution coefficient of a chains-filled container after $i$th collision |
| $V_a$ | Velocity of a container after collision |
| $V_b$ | Velocity of a container before collision |
| $n_b$ | Numbers of rebound |
| $\tau^*$ | Time between the initial free fall and first rebound |
| $\tau_0$ | Total flight time of an empty container |
| $\tau$ | Total flight time duration of a chains-filled container |
| $h$ | Vertical position of the geometric center of a container |
| $H_0$ | Initial height of a container |

## Acknowledgements

The work was supported by the Chinese National Science Foundation, Project Nos. 11104013 and 11475018.

## Appendix A. Supplementary data

Supplementary data to this article can be found online at http://dx.doi.org/10.1016/j.powtec.2016.04.019.

## References

[1] H.V. Panossian, Structural damping enhancement via nonobstructive particle damping technique, J. Vib. Acoust. 114 (1992) 101–105.
[2] M.M. Sadek, B. Mills, The Application of the Impact Damper to the Control of Machine Tool Chatter, Proceedings of the Seventh International Machine Tool and Die Research Conference 1966, pp. 243–257.
[3] R. Kielb, F.G. Macri, D. Oeth, A.D. Nashif, P. Macioce, H. Panossian, F.L. Lieghley, Advanced Damping Systems for Fan and Compressor Blisks, Proceedings of the 4th National Turbine Engine High Cycle Fatigue Conference, Monterey, CA, 1999.
[4] S.S. Simonian, Particle beam damper, SPIE 2445 (1995) 149–160.

C. Xu et al. / Powder Technology 297 (2016) 367–373 373

[5] S. Ashley, A new racket shakes up tennis, Mech. Eng. 117 (1995) 80–81.
[6] M. Heckel, A. Sack, J.E. Kollmer, T. Pöschel, Granular dampers for the reduction of vibrations of an oscillatory saw, Phys. A 391 (2012) 4442–4447.
[7] M. Sánchez, L.A. Pugnaloni, Effective mass overshoot in single degree of freedom mechanical systems with a particle damper, J. Vib. Acoust. 330 (2011) 5812–5819.
[8] A. Sack, M. Heckel, J.E. Kollmer, F. Zimber, T. Pöschel, Energy dissipation in driven granular matter in the absence of gravity, Phys. Rev. Lett. 111 (2013) 018001.
[9] A. Sack, M. Heckel, J.E. Kollmer, T. Pöschel, Probing the validity of an effective-one-particle description of granular dampers in microgravity, Granul. Matter 17 (2015) 73–82.
[10] X.M. Bai, L.M. Keer, Q.J. Wang, R.Q. Snurr, Investigation of particle damping mechanism via particle dynamics simulations, Granul. Matter 11 (2009) 417–429.
[11] K.M. Mao, M.Y. Wang, Z.W. Xu, T.N. Chen, Simulation and characterization of particle damping in transient vibrations, J. Vib. Acoust. 126 (2004) 202–211.
[12] K.M. Mao, M.Y. Wang, Z.W. Xu, T.N. Chen, DEM simulation of particle damping, Powder Technol. 142 (2004) 154–165.
[13] L.N. Zou, X. Cheng, M.L. Rivers, H.M. Jaeger, S.R. Nagel, The packing of granular polymer chains, Science 326 (2009) 408.
[14] P.P. Wen, N. Zheng, L.S. Li, H. Li, G. Sun, Q.F. Shi, Polymerlike statistical characterization of two-dimensional granular chains, Phys. Rev. E 85 (2012), 031301.
[15] K. Safford, Y. Kantor, M. Kardar, A. Kudrolli, Structure and dynamics of vibrated granular chains: comparison to equilibrium polymers, Phys. Rev. E 79 (2009), 061304.
[16] E. Brown, A. Nasto, A.G. Athanassiadis, H.M. Jaeger, Strain stiffening in random packings of entangled granular chains, Phys. Rev. Lett. 108 (2012), 108302.
[17] P.P. Wen, G. Wang, D. Hao, N. Zheng, L. Li, Q. Shi, Bottom stresses of static packing of granular chains, Phys. A 419 (2015) 457–463.
[18] M. Hu, Q.S. Mu, N. Luo, G. Li, N.B. Peng, Behavior of hollow balls containing granules bouncing repeatedly off the ground, EPL 103 (2013), 14003.
[19] F. Pacheco-Vázquez, S. Dorbolo, Rebound of a confined granular material: combination of a bouncing ball and a granular damper, Sci. Rep. 3 (28) (2013) 2158.
[20] F. Pacheco-Vázquez, F. Ludewig, S. Dorbolo, Dynamics of a grain-filled ball on a vibrating plate, Phys. Rev. Lett. 113 (2014), 118001.